\documentclass[12pt]{article}
\usepackage{amssymb}
\usepackage{amsmath}
\usepackage{epsfig}
\usepackage{dcolumn}
\usepackage{rotating}
\usepackage{color}
\usepackage{slashed}
\definecolor{rosso}{cmyk}{0,1,1,0.4}
\definecolor{rossos}{cmyk}{0,1,1,0.55}
\definecolor{rossoc}{cmyk}{0,1,1,0.2}
\definecolor{blu}{cmyk}{1,1,0,0.3}
\definecolor{blus}{cmyk}{1,1,0,0.6}
\definecolor{bluc}{cmyk}{1,1,0,0.1}
\definecolor{verde}{cmyk}{0.92,0,0.59,0.25}
\definecolor{verdec}{cmyk}{0.92,0,0.59,0.15}
\definecolor{verdes}{cmyk}{0.92,0,0.59,0.7}

\newcommand{\ba}{\begin{eqnarray}}
\newcommand{\ea}{\end{eqnarray}}
\newcommand{\be}{\begin{equation}}
\newcommand{\ee}{\end{equation}}
\newcommand{\bi}{\begin{itemize}}
\newcommand{\ei}{\end{itemize}}
\newcommand{\al}{\alpha}

\newcommand{\ga}{\gamma}

\newcommand{\da}{\delta}
\newcommand{\la}{\lambda}

\newcommand{\za}{\zeta}

\newcommand{\en}{\epsilon}
\newcommand{\oa}{\omega}
\newcommand{\Ga}{\Gamma}

\newcommand{\cD}{{\cal D}}

\newcommand{\cO}{{\cal O}}

\newcommand{\cK}{{\cal K}}
\newcommand{\cM}{{\cal M}}

\newcommand{\e}{\overline}

\newcommand{\p}{\partial}

\newcommand{\n}{\nabla}

\newcommand{\ra}{\rightarrow}

\newcommand{\LF}{\left(}
\newcommand{\RF}{\right)}
\newcommand{\LT}{\left[}
\newcommand{\RT}{\right]}
\newcommand{\Ld}{\left.}
\newcommand{\Rd}{\right.}

\newcommand{\me}{\e{m}}

\newcommand{\2}{\frac{1}{2}}

\newcommand{\4}{\frac{1}{4}}

\newcommand{\mx}{\mbox}
\newcommand{\mt}{\mathtt}
\newcommand{\mtf}{\mathtt{f}}
\newcommand{\mand}{\mx{ and }}

\newcommand{\vs}{\vspace{5mm}\\}
\newcommand{\non}{\nonumber\\}

\begin{document}
\tolerance=100000
\thispagestyle{empty}
\vspace{1cm}

\begin{center}
{\LARGE \bf
Towards LHC Physics with Non-local Standard Model
 \\ [0.15cm]
}
\vskip 2cm
{\large Tirthabir Biswas\footnote{tbiswas@loyno.edu}}$^a$ and
{\large Nobuchika Okada\footnote{okadan@ua.edu}}$^{b}$,

\vskip 7mm
{\it $^a$  Department of Physics, Loyola University, \\ 6363 St. Charles Avenue, Box 92\\
New Orleans, LA 70118, USA} \\
\vskip 3mm
{\it $^b$  Department of Physics and Astronomy, University of Alabama, \\  Tuscaloosa, AL 35487-0324, USA
} \\
\vskip 3mm
\end{center}
\date{\today}

\begin{abstract}
We take a few steps towards constructing a string-inspired nonlocal extension of the Standard Model. We start by illustrating how quantum loop calculations can be performed in nonlocal scalar field theory. In particular, we show the potential to address the hierarchy problem in the nonlocal framework.
Next, we construct a nonlocal abelian gauge model and derive modifications of the gauge interaction vertex and field propagators.
We apply the modifications to a toy version of the nonlocal Standard Model and investigate collider phenomenology.
We find the lower bound on the scale of non-locality from the 8 TeV LHC data to be $2.5-3$ TeV
\end{abstract}

\newpage

\setcounter{page}{1}

\tableofcontents
\section{Introduction}
Strings, by their very definition, are nonlocal objects: even classically they do not interact with each other at a specific spatial point, but rather over a region in space. Not surprisingly, non-local structures are a recurrent theme in stringy theories/models. For instance, this is the case in string field theory (SFT)~\cite{sft1,sft2,sft3} and various toy models of string theory such as $p$-adic strings~\cite{padic1,padic2,padic3,Frampton-padic}, zeta strings~\cite{zeta}, and strings quantized on a random lattice~\cite{random1,random2,random3,ghoshal}. More general non-local theories that do not respect Lorentz symmetry arise in noncommutative field theories~\cite{ncft1,ncft2}, field theories with a minimal length scale~\cite{minimal} (such as doubly special relativity), fluid dynamics and quantum algebras~\cite{kdv}. Many theorists have used these theories  to probe different stringy phenomena such as  tachyon dynamics~\cite{tachyon,sen-tachyon,zwiebach}, excitations on branes~\cite{minahan}, Regge behavior~\cite{marc}, stringy solitons~\cite{zwiebach,solitons,BCK-solitons}, and stringy thermodynamics~\cite{BCK-prl,BCK,Bluhm, BKR}. In particular, by applying  finite temperature field theory techniques several remarkable results were obtained.  The $p$-adic partition function, at least at the 2-loop level, exhibited thermal duality~\cite{BCK-prl}: $Z(T)\sim Z(M^2/4\pi T)$, which was expected from stringy arguments~\cite{Atick-witten}, but had not been explicitly demonstrated beforehand. The partition function also reproduced known features of the so called ``Hagedorn Phase''~\cite{Deo,Vafa-Tseytlin}, and provided new insights. Inspired by stringy nonlocal physics in this paper we adopt a phenomenological approach and take a few steps towards building a specific nonlocal extension of the Standard Model, and also investigate potential signatures of these theories at the Large Hadron Collider or LHC.

A  Poincare invariant  nonlocal formulation of the standard model was provided in~\cite{moffat-prd}, see also~\cite{moffat-qft}.  
While~\cite{moffat-prd,moffat-qft} considers nonlocal gauge transformations, in our model the gauge transformations remain local, it is the interactions that become non-local as was first discussed in~\cite{efimov-qed1,efimov-qed2}. Finally, let us also point out that  similar string-inspired nonlocal interactions was also studied before in~\cite{string_LHC}, but the modifications were introduced in a somewhat ad-hoc fashion rather than deriving it from an action as we will in this paper.


Now, apart from the fact that the nonlocal modifications in the form of an exponential damping of the propagators are inspired by string theory, arguably the strongest candidate for a unified theory of gravity and particle physics, these theories have several attractive features. Unlike most other  higher derivative theories, the nonlocal higher derivative models are expected to be free from ghosts~\cite{Efimov-unitarity,moffat-qft}, at least at the perturbative level, and are also thought to  have a well-posed initial value problem \cite{zwiebach,Volovich-math,Vladimirov-math,Vladimirov-math2,neil-kamran,gianluca-math}. What makes them particularly exciting is that the propagator in these nonlocal theories are exponentially damped at high momentum often making all the scattering amplitudes finite. Thus these theories are strong candidates for a truly UV-complete field theory.  The exponential damping can be interpreted as a smooth Lorentz-invariant way of incorporating the fact that at the smallest distance scales one does not expect a space-time continuum. Therefore arbitrary small spatio-temporal fluctuations  are ``unphysical'' and needs to be eliminated. The traditional approach is via regularization/renormalization. In contrast, in nonlocal theories the  mass scale, $M$, associated with the exponential damping of the propagator behaves like a physical parameter (in String theory it would be related to the string tension) eliminating smaller than $M^{-1}$ length scale fluctuations. This leads to small differences in the scattering cross-sections between the  the local field theories and their nonlocal counterparts. These are the kind of differences that we are going to try to estimate in the context of LHC measurements.

It is important, however, to point out that even if these infinite-derivative theories do not end up describing the   interactions  at the most fundamental level, they could still be ``effectively'' capturing dynamics of the more fundamental theory up to a certain scale. For instance, such kinetic modifications have been very successfully used in nuclear physics~\cite{nuclear-nonlocal}. In this paper, we take this more modest approach where we look at higher derivative modifications to the kinetic part of the Standard Model action that can encapture nonlocal physics. Indeed, this is similar in spirit to the higher dimensional operators that is widely used in the literature, see for instance~\cite{higher-operators}, to parameterize new physics beyond the Standard Model.  It is also worth pointing out that previously, the nonlocalization in such theories was  studied  as a regularization scheme~\cite{polchinski,moffat-woodard,woodard} and was found to have some distinct advantages over some of the more traditional schemes such as dimensional regularization, $\za$-function regularization and Pauli-Villars method.

Another motivation to studying such nonlocal effects in SM  comes from trying to solve the ``hierarchy'' problem, as was initially pointed out in~\cite{moffat-prd}. Just as a supersymmetry breaking scale, $M_{\mt{susy}}= \cO(\mt{TeV})$ can keep the  mass of the Higgs stay small by ensuring the bosonic and fermionic loop contributions  cancel beyond $M_{\mt{susy}}$, the exponential suppression of all spatial fluctuations shorter than $M^{-1}$ in the nonlocal models can produce the same result if $M= \cO(\mt{TeV})$. This provides us with an additional motivation to study scenarios where $M = \cO(\mt{TeV})$. Obviously, a complete resolution of the hierarchy problem  also requires a mechanism that can explain why M  is 15 orders of magnitude smaller than the Planck scale; the nonlocal interaction can only prevent quantum mass corrections from diverging, if a TeV nonlocal scale is present in the first place. However, this is still an interesting advantage of the nonlocal theories and in this regard the situation is no worse than SUSY~\footnote{In this context it is worth noting that while usually the string scale is thought to be only a few orders of magnitude smaller than the Planck scale, there are several stringy compactification schemes where the string scale can be made much lower. Different mechanisms have been proposed to explain the hierarchy between the Planck and the electro-weak scale by postulating that the Planck scale is not a fundamental but rather a derived scale which happen to obtain a large value due to warping~\cite{RS1,RS2} or cosmological evolution in the ``early'' universe~\cite{BN-hierarchy}}.

In this  paper our first goal is to construct a nonlocal version of the abelian sector of the Standard Model  taking care to preserve gauge invariance. Next we will compute the differences in cross sections between the Nonlocal Standard Model (NLSM) and SM at high energy colliders and potential signatures of these models that may be visible in the next LHC run.

Our paper is organized as follows: In section~\ref{sec:scalar}, with the purpose of illustration we introduce the string-inspired nonlocal modification in a scalar toy model. We discuss the renormalization prescription in these theories and show how it leads to small differences from the usual field theoretic results in scattering cross sections. Next in section~\ref{sec:NLSM}, we construct a nonlocal version of the abelian sector of the SM which preserves gauge invariance. We also obtain the  Feynman rules for this model. In section~\ref{sec:LHC}, we compute  some of the cross sections in NLSM that is relevant to high energy collider experiments. We obtain lower bounds on the parameters of NLSM from the current experimental data.
In section~\ref{sec:conclusions}, we summarize our results and discuss future directions.
\section{Nonlocal Scalar Field Theory}
\label{sec:scalar}
\subsection{Action}
Canonical examples of nonlocal actions that appear in string literature can be written as
\be
S=\int d^Dx \LT\2\phi \cK(\Box)\phi -V_{\mt{int}}(\phi)\RT \ ,
\label{nlaction}
\ee
in the ``$+---$'' metric signature convention that we will employ throughout this manuscript. The kinetic operator $\cK(\Box)$ contains an infinite series of higher derivative terms.
For instance, $\cK(\Box)=-e^{\Box/M^2}$ for stringy toy models based on $p$-adic numbers~\cite{padic1,padic2,padic3} or random lattices~\cite{random1,random2,random3,ghoshal,marc}, and $\cK(\Box)=-(\Box+m^2)e^{\Box/M^2}$  in String Field theory~\cite{sft1,sft2,sft3} (for a review see~\cite{sft-review}), where $m^2(<0)$ and $M^2(>0)$ are proportional to the string tension. Here we are going to take a phenomenological approach and investigate particle physics implications if $M=\cO(\mt{TeV})$. As mentioned in the introduction, the first motivation comes from the fact that although these theories contain higher derivatives, they do not contain ghosts (at least perturbatively) and are able to retain the improved UV  behavior expected in higher derivative theories. To see this explicitly, one can consider a fourth order scalar theory with $\cK(\Box)=-\Box(1+{\Box\over m^2})$. The corresponding propagator reads
\be
\Pi(p^2)\sim{-m^2\over p^2(p^2 -m^2)}\sim{-1\over p^2 -m^2}+{1\over p^2}
\ee
From the pole structure of the propagator it is clear that the theory contains two physical states, but unfortunately the massive state has the ``wrong'' sign for the residue indicating that it is a ghost. Once interactions are included, it makes the classical theory unstable, and the quantum theory non-unitary. The stringy kinetic modifications, on the other hand, combine to be an exponential which is an entire function without any zeroes. In other words, it does not introduce any new states, ghosts or otherwise, and only ameliorates the UV behavior with an exponential suppression.

The second motivation, as was also alluded to in the introduction, has to do with the fact that in these nonlocal theories corrections to the mass of a scalar field typically goes as $\sim M^2$. Therefore, such modifications provide a way to ameliorate the hierarchy problem if $M= \cO(\mt{TeV})$. To see this explicitly, let us start with  a  simple nonlocal $\la\phi^4$ theory of scalar fields:
\be
S=\int d^4x\ \LT-\2\phi e^{{\Box+m^2\over M^2}}(\Box + m^2)\phi -{\la\over 4!}\phi^{4}\RT
\ee
Here  the normalization of $\phi$ is so chosen that the residue at the $p^2=m^2$ pole is unity.
In Eucleadian space ($p^0\ra ip^0$) the propagator is given by
\be
\Pi(p^2)=-{ie^{-{p^2+m^2\over M^2}}\over p^2+m^2}
\ee
while the vertex factor is, as usual, given by $-i\la$.
\subsection{Two point function \& the hierarchy problem}
To understand how the quantum loop computations work, let us first compute the one loop 2-pt function with zero external momentum.
We have~\footnote{In going from Minkowski to Eucleadian integral we also have to replace $d^4p$ with $id^4p$.}
\be
\Ga_2=-{i\la\over 2}\int {d^4k\over (2\pi)^4}\ {e^{-{(k^2+m^2)\over M^2}}\over k^2+m^2}=-i{\la e^{-{m^2\over M^2}}\over 16\pi^2}\int dk\ {k^3 e^{-{k^2\over M^2}}\over k^2+m^2}
\ee
The above integral is finite and has an analytical expression
\be
\Ga_2=-i{\la M^2 e^{-{m^2\over M^2}}\over 32\pi^2}\left[1 +e^{{m^2\over M^2}}\LF{m^2\over M^2}\RF  Ei \left( -\frac{m^2}{M^2} \right)  \right]
\ee
where $Ei$ is the exponential-integral function defined via
\be
Ei(z)=-\int^{\infty}_{-z} {e^{-t}\over t} dt
\ee
where the Cauchy Principal Value is taken.

The mass correction is naively given by
\be
\da m^2=i\Ga_2={\la \over 32\pi^2}\left[e^{-{m^2\over M^2}} +\LF{m^2\over M^2}\RF  Ei \left( -\frac{m^2}{M^2} \right)  \right]M^2
\label{deltam2}
\ee
The $Ei$ function has a mild divergence as $z\ra 0$, but $z Ei(z)\ra 0$ as $z\ra 0$. Thus we see that when $M\gg m$
\be
\da m^2={\la \over 32\pi^2}M^2
\label{deltamsquare}
\ee
In other words, the ``mass correction'' grows linearly with $M$. Thus if $M= \cO(\mt{TeV})$, the model can address the hierarchy problem as was discussed in~\cite{moffat-prd}, (\ref{deltamsquare}) provides an explicit corroboration. Just like with the SUSY breaking scale, this provides an encouragement to search for nonlocal effects at LHC.

There is a subtle point worth noting: Due to the exponential cutoff in the ``bare'' propagator, $\da m^2$ is not exactly $i\Ga_2$. Resumming all the 1PI diagrams with the bare propagators sandwiched between the 1PI contributions, one obtains the physical propagator, $\Pi_{\mt{phys}}(p^2)$, to be
\be
\Pi_{\mt{phys}}(p^2)=-{ie^{-{p^2+m^2\over M^2}}\over p^2+m^2+i\Ga_2e^{-{p^2+m^2\over M^2}}}
\ee
In other words the 1-loop correction to the mass depends on the momentum, and is given by
\be
\da m^2(p^2)={\la \over 8\pi^2}{M^2\over 2}\left[e^{-{m^2\over M^2}} +\LF{m^2\over M^2}\RF  Ei \left( -\frac{m^2}{M^2} \right)  \right]e^{-{p^2+m^2\over M^2}}
\ee
which reduces to (\ref{deltam2}) at $p^2=-m^2$.
\subsection{Four point scattering amplitude}
It is instructive to compute the 4pt-scattering amplitude in the nonlocal theory and compare  the results with the local field theoretic calculations. In the process we will also see why the quantum loops in these nonlocal theories remain finite and provide small corrections to their local counterparts. Let us start with the one loop four point function, $\Ga_4$ for zero external momenta. This is given by
\be
\Ga_4={i\la^2 e^{-{2m^2\over M^2}}\over 2}\int {d^4k\over (2\pi)^4}\ {e^{-{2k^2\over M^2}}\over (k^2+m^2)^2}={i\la^2 e^{-{2m^2\over M^2}}\over 16\pi^2}\int dk\ {k^3 e^{-{2k^2\over M^2}}\over (k^2+m^2)^2}
\ee
Again, one can compute it exactly:
\be
\Ga_4=-{i\la^2 e^{-{2m^2\over M^2}}\over 32\pi^2}\left[1 +\LF1+{2m^2\over M^2}\RF e^{{2m^2\over M^2}} Ei \left( -\frac{2m^2}{M^2} \right)  \right]
\ee
$\Ga_4$ is indeed finite and is essentially the 1-loop contribution to the effective potential at the quartic level. One observes that  $\Ga_4$ diverges mildly (similar to a logarithm) as $m/M\ra 0$. This is again to be expected from the local field theory results. A few important points however emerge: First, the UV and IR divergences are tied together, the IR divergence is expected as $m\ra 0$, while the UV divergence corresponds to $M\ra \infty$, we now have a single divergent combination. Second, it is clear that since the dependence on our ``cut-off'' scale is mild, it will provide different predictions for the scattering amplitude (as compared to usual renormalization) but the corrections are going to be small.

Let us next look at the scattering amplitude when say two particles with momenta $p_1$ and $p_2$ scatter. The $s$-channel loop integral is given by
\be
A(s)={i\la^2 e^{-{2m^2\over M^2}}\over 2}\int {d^4k\over (2\pi)^4}\ {e^{-{k^2\over M^2}}e^{-{(k+P)^2\over M^2}}\over (k^2+m^2)((P+k)^2+m^2)}
\ee
where $P=p_1+p_2$. There is no analytical expression for the above, but one way to make progress is to introduce Schwinger parameters and rewrite the above integral as
\be
A(s)={i\la^2 e^{-{2m^2\over M^2}}\over 2}\int {d^4k\over (2\pi)^4}\int_0^{\infty} d\xi_1\int_0^{\infty} d\xi_2\ \ e^{-{k^2\over M^2}}e^{-{(k+P)^2\over M^2}} e^{-{(k^2+m^2) \xi_1}}e^{-((k+P)^2+m^2)\xi_2}
\ee
We therefore need to just perform some gaussian integrals. In a straight forward way one finds
\be
A(s)={iM^4\la^2 e^{-{2m^2\over M^2}}\over 32\pi^2}\int_0^{\infty} d\xi_{1}\int_0^{\infty} d\xi_2\ {e^{-m^2(\xi_1+\xi_2)} e^{{s\over M^2}{(1+\xi_1M^2)(1+\xi_2M^2)\over 2 +M^2(\xi_1+\xi_2)}}\over (2 +M^2(\xi_1+\xi_2))^2}
\ee
Here $s=-P^2$ is the usual (Minkowski) Mandelstam variable.

The main point here is that the  scattering cross-section depends on three parameters $m^2,M^2,\lambda$ as opposed to only two parameters in the local field theory, $m^2_r,\la_r$, the $r$ referring to renormalized quantities. This means that in general we will have a different result as compared to conventional renormalization. However, the dependence on $M^2$ is weak, and that is why one reproduces the local field theory results in the large $M$ limit. At this point it is also worth pointing out that interactions that are traditionally considered non-renormalizable produce finite loops once the propagators are made nonlocal. For instance, loops for a $\phi^6/M_6^2$ interaction are finite, and also give rise to a valid perturbative loop expansion as long as $M\lesssim M_6$, see~\cite{marc,BCK} for similar calculations in the context of $p$-adic strings. This suggests that ``nonlocalization'' may be a way to eliminate quantum divergences in nonrenormalizable theories which may have  profound implications for fundamental physics such as gravity, see~\cite{BMS,BGKM,BMT,moffat-prd,Moffat-gravity,Modesto,Tsujikawa-gravity,modesto-gianluca} for efforts in this direction.
\subsection{QFT vs. NLFT}
We want to compare the results in a theory where $M$ is a physical parameter versus the usual renormalization prescription where $M$ effectively behaves as a regulator.

To see this, let us first write down the complete 2-particle scattering amplitude. At the 1-loop level, this is a sum of three diagrams, the $s$, the $t$ and the $u$ channel.The amplitude thus reads
\be
i\cM=-i\la+[A(s)+A(t)+A(u)]
\ee
Changing the integration variable from $\xi\ra M^2\xi$, we have
\be
A(s)={i\la^2 e^{-2\me^2}\over 32\pi^2}\int d\xi_1\int d\xi_2\ {e^{-\me^2(\xi_1+\xi_2)} e^{{s\over M^2}{(1+\xi_1)(1+\xi_2)\over 2 +\xi_1+\xi_2}}\over (2 +\xi_1+\xi_2)^2}
\ee
where $\me\equiv m/M$.

While no analytic solution to the above exist, we can come up with an upper bound for the above integral as follows:
\ba
A(s)&=&{i\la^2 e^{-2\me^2}\over 32\pi^2}\int d\xi_1\int d\xi_2\ \sum_n{e^{-\me^2(\xi_1+\xi_2)} \over (2 +\xi_1+\xi_2)^2}{1\over n!}\LF{{s\over M^2}{(1+\xi_1)(1+\xi_2)\over 2 +\xi_1+\xi_2}}\RF^n\\
&=&{i\la^2 e^{-2\me^2}\over 32\pi^2}\sum_n{1\over n!}\LF{{s\over M^2}}\RF^n\int d\xi_1\int d\xi_2\ {e^{-\me^2(\xi_1+\xi_2)} \over (2 +\xi_1+\xi_2)^2}\LF{{(1+\xi_1)(1+\xi_2)\over 2 +\xi_1+\xi_2}}\RF^n
\label{As}\non
\ea
Let us evaluate the integral
\ba
I_n&\equiv& \int d\xi_1\int d\xi_2\ {e^{-\me^2(\xi_1+\xi_2)} \over (2 +\xi_1+\xi_2)^2}\LF{{(1+\xi_1)(1+\xi_2)\over 2 +\xi_1+\xi_2}}\RF^n<\int d\xi_1\int d\xi_2\ e^{-\me^2(\xi_1+\xi_2)} \LF1+\xi_2\RF^{n-2}\non
&=&-{e^{\me^2}E_{2-n} \left({\me}^{2} \right)\over {\me}^{2}}\ ,
\ea
where
\be
E_n(z)\equiv \int_1^\infty dt\ {e^{-zt}\over t^{n}}\ .
\ee

Since $E_{2-n}  \left({\me}^{2} \right)$ diverges as $\me\ra 0$, only for $n=0,1$, it is clear that all the $n>1$ terms are finite. The first two terms in (\ref{As}) are indeed divergent, but they can be absorbed ``within'' a physical quantity. To see this let us define
\be
\la_p\equiv -\cM(s=4m^2,t=u=0)=\la-{\la^2 e^{-2\me^2}\over 32\pi^2}\LT\int d\xi_1\int d\xi_2\ {e^{-\me^2(\xi_1+\xi_2)} \LF 2+e^{4\me^2{(1+\xi_1)(1+\xi_2)\over 2 +\xi_1+\xi_2}}\RF\over (2 +\xi_1+\xi_2)^2}\RT
\ee
Expanding the exponential  we find that the first two terms diverge:
\be
\la_p=\la-{\la^2 e^{-2\me^2}\over 32\pi^2}\LT\int d\xi_1\int d\xi_2\ e^{-\me^2(\xi_1+\xi_2)} \LF{ 3+4\me^2{(1+\xi_1)(1+\xi_2)\over 2 +\xi_1+\xi_2}\over (2 +\xi_1+\xi_2)^2}+ \cO(\me^4)\RF\RT
\ee
However, this is precisely the divergence that one finds in the first two terms in the expression for the scattering amplitude
\ba
-\cM(s,t,u)&=&\la-{\la^2 e^{-2\me^2}\over 32\pi^2}\LT\int d\xi_1\int d\xi_2\ {e^{-\me^2(\xi_1+\xi_2)} \LF 3-{4m^2\over M^2}{(1+\xi_1)(1+\xi_2)\over 2 +\xi_1+\xi_2}\RF\over (2 +\xi_1+\xi_2)^2}\RT\non
&+& \cO\LF{s\over M^2},{t\over M^2},{u\over M^2}\RF
\ea
by virtue of the identity $s+t+u=4m^2$. It is therefore clear that one can rewrite the scattering amplitude as
\be
-\cM(s,t,u)=\la_p+(\mx{ terms regular as } M\ra \infty)
\ee
This means that for large values of $M$ the $\cM$ is not going to grow with $M$, which would have made these theories untenable phenomenologically, but rather the corrections are going to be suppressed with $M$. This ensures that we will only get small nonlocal corrections to the usual field theory results.

So how are the renormalizable theories different from what we are proposing? In the standard renormalization prescription $M$ (or any other regulator for that matter) is not a physical parameter but only used to keep track of the infinities. Thus, once the inifinities are absorbed within the physical quantities, one takes appropriate limits of the regulators, $M\ra \infty$ in this case, to obtain results which no longer depend on the regulators. In contrast, if $M$ is a physical parameter, then the non-divergent pieces retains it's dependence on the regulator, $M$. If $M$ is much too large as compared to the other mass scales involved ($m, s,t,u$), then the difference between the ``physical'' and ``traditional'' approach to renomalization will be negligible. However, if for instance, $M= \cO(\mt{TeV})$, then we will start observing deviations between the two approaches in the current/future collider experiments.

To see this most explicitly, let us look at the $n=2$ terms. We have
\ba
-\cM-\la_p&=&-{\la^2 e^{-2\me^2}\over 32\pi^2}\LF{s^2+t^2+u^2-16m^4\over 2M^4}\RF\int d\xi_1\int d\xi_2\ {e^{-\me^2(\xi_1+\xi_2)} (1+\xi_1)^2(1+\xi_2)^2\over (2 +\xi_1+\xi_2)^4}\non
&=&-{\la^2 e^{-2\me^2}\over 128\times 15\pi^2}\LF{s^2+t^2+u^2-16m^4\over 2m^4}\RF\LF 4+8{\me}^{2}-{27}\,{\me}^{4}-{28}\,{\me}^{6}-{4}\,{\me}^{8}\Rd\non
&-&
{\me}^{6}\Ld{{\rm e}^{2\,{\me}^{2}}}{\it Ei} \left(-2{\me}^{2} \right) \LF 80+60{\me}^{2}+8{\me}^{4}\ .
 \RF\RF
\ea
Clearly, all the terms are finite when $\me\ra 0$. According to the usual renormalization prescription, one takes the limit $M\ra \infty$ or $\me\ra0$. This would give us
\be
-\cM_{\mt{loc}}=\la_p-{\la_p^2\over 32\times 15\pi^2}\LF{s^2+t^2+u^2-16m^4\over 2m^4}\RF+\cO(\la_p^3)
\ee
In our approach, however, we will have corrections:
\be
-\cM=\la_p-{\la_p^2\over 32\times 15\pi^2}\LF{s^2+t^2+u^2-16m^4\over 2m^4}\RF\LT 1-{43\over 4}\me^4+\cO(\me^6\ln\me)\RT+\dots
\ee
Thus we  see that while the nonlocal corrections to local field theory amplitudes are expected to be small, nevertheless they may be detectable!

\section{Nonlocal Abelian Gauge Theory}
\label{sec:NLSM}
\subsection{Pure Gauge}
Standard Model, of course, is a gauge theory and our first task in constructing a nonlocal version of SM is, therefore, to ``nonlocalize'' gauge theories. In~\cite{moffat-prd,moffat-woodard} this was achieved by advocating a nonlocal gauge transformation. Here we are going to pursue a more direct approach which is more analogous to the way one introduces nonlocality in the scalar field action, except that we will use covariant derivatives in the exponential kinetic operator instead of normal derivatives. This procedure makes the action manifestly gauge invariant under the usual local gauge transformation, but the interactions become nonlocal. This has the  disadvantage that at 1-loop the quantum amplitudes do not all remain finite and therefore will require implementing the usual regularization/renormalization procedure~\footnote{The most important corrections that are relevant for LHC appears at the tree-level and therefore we don't need to worry about loop graphs for phenomenological purposes.}. This is not really a serious issue when it comes to gauge theories which are renormalizable, but it could be an important consideration if one attempts to construct a nonlocal quantum theory of gravity. Indeed, the nonlocal theories developed along the lines of~\cite{moffat-prd,moffat-woodard,Moffat-gravity} might be able to avoid this problem outright, but recent investigations in trying to understand the quantum divergences in nonlocal gravity theories of the form we are discussing here indicate that while 1-loop diagrams can diverge, once they are renormalized, the higher loops become finite~\footnote{For exponential kinetic theories, what essentially determines whether a diagram diverges or not depends on the sign of the exponent that appears in front of the momentum squares. For gauge/gravity theories discussed in this paper, while the propagators are exponentially suppressed, the vertices are enhanced approximately by the same factor, see (\ref{gauge-vertex}) for instance. Thus if $P$ and $V$ represents the number of propagators and vertices, the sign is determined by $V-P=1-L$ according to topological identity, where $L$ is the number of loops, see~\cite{BMT,Modesto} for details. This means that while we expect the $L=1$ graphs to be divergent, $L>1$ graphs should be convergent! This is what current investigations~\cite{BMT} seem to suggest, although several subtle effects, including the importance of using the dressed propagator as opposed to the bare propagator, still require a comprehensive investigation.}. Finally, while our procedure can be extended to nonabelian gauge theories, that is technically much more challenging to implement as compared to the abelian gauge theories.  Accordingly, in this paper we are going to limit ourselves to the latter.

For abelian theories, since the field strength,
\be
F_{\mu\nu}=\p_{[\mu}A_{\nu]}\ ,
\ee
 itself is gauge invariant, the implementation of the nonlocal modification is rather straightforward. The action,
\be
S_{\mt{g}}=-\4\int d^4x\ F^{\mu\nu}e^{-{\Box\over M_g^2}} F_{\mu\nu}
\ee
is trivially gauge invariant. As usual, we need to supplement this with a gauge fixing procedure. As in the standard QED, one can introduce a gauge fixing function, $G(A)=g(\Box)\p_{\mu}A^{\mu}$, via the delta function
\be
\int \cD\al(x)\da(G(A^{\al}))\mt{det}\LF{\da G(A^{\al})\over \da \al}\RF=1\ ,
\ee
in the path integral
\be
Z\equiv \int \cD A\ e^{iS[A]}\ .
\ee
Only now we have to choose a higher derivative gauge fixing function of the form
\be
G(A)=g(\Box)\p_{\mu}A^{\mu}\ .
\ee
This procedure  was, in fact, outlined previously in~\cite{Stelle} while discussing higher derivative gravity theories. The main point is that since the gauge transformed field,
\be
A^{\al}_{\mu}\equiv A_{\mu}+{1\over e}\p_{\mu}\al(x)\ ,
\ee
is only linear in $\al$, it's derivative with respect to $\al$ is a constant. Thus, as in usual QED,  the determinant in our nonlocal theory is a also constant and can be taken out of the path integral:
\be
Z\equiv \mt{det}\LF{\da G(A^{\al})\over \da \al}\RF\int \cD\al\int \cD A\ e^{iS[A]}\da(G(A^{\al}))\ .
\ee
Following the usual gauge fixing procedure, since $\cD A^{\al}=\cD A$ and $S[A^{\al}]=S[A]$, we obtain
\be
Z=\mt{det}\LF{\da G(A^{\al})\over \da \al}\RF\int \cD\al\int \cD A\ e^{iS[A]}\da(G(A))\ .
\ee
Now noting that the same procedure holds for a general class of gauge fixing function
\be
G(A)=g(\Box)\p_{\mu}A^{\mu}-\oa(x)\ ,
\ee
and then integrating over $\oa(x)$ using a gaussian weight we obtain the gauge fixed path integral as
\be
Z\ra N(\xi)\mt{det}\LF{g(\Box)\Box\over e}\RF\LF\int \cD\al\RF\int \cD A\ e^{iS[A]}e^{-i\int d^4x\ G(A)^2\over 2\xi}
\ee
The first three terms cancel out while calculating any scattering process when properly normalized, and therefore, we have effectively a new gauge fixed action given by
\be
S=\int d^4x\ \LT-\4 F^{\mu\nu}e^{-{\Box\over M_g^2}} F_{\mu\nu} -{[g(\Box)\p_{\mu}A^{\mu}]^2\over 2\xi}\RT
\ee

Appropriate integration by parts lead us to the more convenient form
\be
S=\2\int d^4x\ \LT A^{\mu}e^{-{\Box\over M_g^2}} (\Box\eta^{\mu\nu}-\p^{\mu}\p^{\nu})A_{\nu} +{1\over \xi}A^{\mu}g^2(\Box)\p^{\mu}\p^{\nu}A_{\nu}\RT
\ee
At this point, it becomes appropriate to choose $g(\Box)=e^{-{\Box\over 2M_g^2}}$
with the result that the QED propagator only gets a nonlocal modification:
\be
\Pi_g(p^2)={-i\eta_{\mu\nu}e^{-{p^2\over M_g^2}}\over p^2+i\en}
\label{NL-propagator}
\ee
in the Feynman gauge $\xi=1$.
\subsection{Including Fermions}\label{sec:fermions}
Our next step towards constructing a nonlocal SM is to include fermions in the story. The most straight forward way to introduce the effect of nonlocality in the fermionic action would be to just add the nonlocal higher derivative terms in the free part of the action, as with the scalars:
\be
S=\int d^4x\ \bar{\psi}_i e^{-{\Box\over M_\mtf^2}}(i\slashed{\p} -m_i)\psi_i
\ee
where $i$ represents the different fermion species~\footnote{In principle different fermion species could have different nonlocal scales, but for simplicity, here we are considering a single nonlocal mass scale for the fermions.}.

Unfortunately, the above nonlocal action is not gauge invariant. Thus, we have to work with the covariant derivatives:
\be
\n_{\mu}=\p_{\mu}+ie A_{\mu}\mand \n^2=\Box+ie (\p\cdot A)+ieA\cdot\p-e^2A^2
\ee
We can now nonlocalize the fermionic action while preserving gauge invariance:
\be
S_{\mt{f}}=\2\int d^4x\ [i\bar{\psi}_ie^{-{\n^2\over M_\mtf^2}}\slashed{\n}\psi_i +h.c.]\equiv S_{\psi}+S_{\psi A}\ ,
\ee
where $S_{\psi}$ contains the free part of the action independent of the gauge fields, while $S_{\psi A}$ contains interaction terms involving the gauge and the fermionic fields.
Let us point out that the first term in the action is not real because the covariant d'Alembertian is non-hermitian, and thus one has to include the hermitian conjugate terms separately. The main complication with the above action arises when one tries to expand the exponential operator and the derivatives can chose to act on $A_{\mu}$ or not with various permutations and combinations possible. Fortunately, one can still rely on perturbative expansions in the fine structure constant, and therefore as a first approximation we can only keep terms that are linear in $A_{\mu}$.

Thus we have
\be
S_{\mt{f}}\approx {i\over 2}\sum_n\int d^4x\ \bar{\psi}_i {(-)^n\n^{2n}\over M_\mtf^{2n}n!}\slashed{\p}\psi_i+ {i\over 2}\int d^4x\ \bar{\psi}_ie^{-{\Box\over M_\mtf^2}}(ie\slashed A\psi_i)+h.c.+\cO(A^2)
\ee
Let us start with
$$
\int d^4x\ \bar{\psi}_i \n^{2n}\slashed{\p}\psi_i\approx\int d^4x\ \bar{\psi}_i \Box^{n}\slashed{\p}\psi_i+ie\sum_{m=0}^{n-1}\int d^4x\ \bar{\psi}_i \Box^{m}( \p\cdot A+A\cdot\p)\Box^{n-m-1}\slashed{\p}\psi_i$$
$$\approx\int d^4x\ \bar{\psi}_i \Box^{n}\slashed{\p}\psi_i+ie\sum_{m=0}^{n-1}\int d^4x\ (\Box^{m}\bar{\psi}_i )( \p\cdot A+A\cdot\p)\Box^{n-m-1}\slashed{\p}\psi_i
$$
Thus putting everything together we have
\be
S_{\mt{f}}\approx -{e\over 2}\sum_n{(-)^n\over M_\mtf^{2n}n!}\sum_{m=0}^{n-1}\int d^4x\ (\Box^{m}\bar{\psi}_i )( \p\cdot A+A\cdot\p)\Box^{n-m-1}\slashed{\p}\psi_i+ {i\over 2}\int d^4x\ \bar{\psi}_ie^{-{\Box\over M_\mtf^2}}\slashed{\n}\psi_i +h.c.
\ee
\subsection{Feynman Rules}
The propagator for the fermions are easy to obtain, they are just the usual ones modulated by the nonlocal factor:
\be
\Pi_f(p^{\mu})={ie^{-{p^2\over M_\mtf^2}}\slashed{p}\over p^2+i\en}
\ee

For the interaction term the relevant action is
\be
S_{\psi A}\approx -{e\over 2}\LT\sum_n {(-)^n\over M_\mtf^{2n}n!}\sum_{m=0}^{n-1}\int d^4x\ (\Box^{m}\bar{\psi}_i )( \p\cdot A+A\cdot\p)\Box^{n-m-1}\slashed{\p}\psi_i+ \int d^4x\ \bar{\psi}_ie^{-{\Box\over M_\mtf^2}}\slashed{A}\psi_i+h.c.\RT
\ee
At this point it is useful to keep track of the hermitian conjugates. Let us start with the last term first:
\be
S_{\psi A, 2}\equiv -{e\over 2}\LT \int d^4x\ \bar{\psi}_ie^{-{\Box\over M_\mtf^2}}(\slashed{A}\psi_i)\RT+h.c.= -{e\over 2}\LT \int d^4x\ (e^{-{\Box\over M_\mtf^2}}\bar{\psi}_i)(\slashed{A}\psi_i)+\bar{\psi}_i\slashed{A}(e^{-{\Box\over M_\mtf^2}}\psi_i)\RT
\ee
The corresponding Feynman vertex function reads
\be
V_2(k_1,k_2)=-{ie\over 2} \LF e^{{k_1^2\over M_\mtf^2}}+e^{{k_2^2\over M_\mtf^2}}\RF\ga^{\mu}
\ee

Next, let us look at the vertex factor coming from the first term:
\be
V_1(k_1,k_2)=-{ie\over 2}\sum_n {(k_{2\mu}+q_{\mu}) \slashed{k}_2\over M_\mtf^{2n}n!}\sum_{m=0}^{n-1}k_1^{2m}  k_2^{2(n-m-1)}=-{ie\over 2}\sum_n {k_{1\mu} \slashed{k}_2\over M_\mtf^{2n}n!}\sum_{m=0}^{n-1}k_1^{2m}  k_2^{2(n-m-1)}
\ee
where $q_{\mu}$ is the photon momentum.

One can now re-sum both the summations:
$$\sum_n {1\over M_\mtf^{2n}n!}\sum_{m=0}^{n-1}k_1^{2m}  k_2^{2(n-m-1)}=\sum_n {1\over M_\mtf^{2n}n!}\LF{k_1^{2n}-k_2^{2m}\over k_1^2-k_2^2}\RF={e^{k_1^2\over M_\mtf^2}-e^{k_2^2\over M_\mtf^2}\over  k_1^2-k_2^2}
$$
to  obtain
\be
V_1(k_1,k_2)=-{ie\over 2}  k_{1\mu} \slashed{k}_2\LF {e^{k_1^2\over M_\mtf^2}-e^{k_2^2\over M_\mtf^2}\over  k_1^2-k_2^2}\RF
\ee

The corresponding hermitian conjugate term reads
\be
S_{2,h.c.}\approx -{e\over 2}\LT\sum_n {(-)^n\over M_\mtf^{2n}n!}\sum_{m=0}^{n-1}\int d^4x\ ( \p.A+A.\p)(\Box^{m}\slashed{\p}\bar{\psi}_i )\Box^{n-m-1}\psi_i\RT
\ee
leading to an overall symmetrization with respect to $k_1,k_2$. The final vertex function reads
\be
V(k_1,k_2)=-{ie\over 2}\LT  (k_{1\mu} \slashed{k}_2+k_{2\mu} \slashed{k}_1)\LF
{e^{k_1^2\over M_\mtf^2}-e^{k_2^2\over M_\mtf^2}\over  k_1^2-k_2^2}\RF+ \LF
e^{{k_1^2\over M_\mtf^2}}+e^{{k_2^2\over M_\mtf^2}}\RF\ga^{\mu} \RT
\label{gauge-vertex}
\ee
In the  $k_1,k_2\ll M_\mtf$ limit, the first term vanishes as it is $\cO(k^2/M^2_\mtf)$, and the second term  reduces to the usual  expression of the QED vertex, as expected.
\section{Collider Phenomenology of the Nonlocal Standard Model}
\label{sec:LHC}
In the previous sections, we have developed the formalism of nonlocal (abelian) gauge theories.
For the completion of the program, we need further developments of the formalism
 to implement non-abelian gauge theories and the Higgs mechanism.
We leave it for future work, but in this section, we discuss potential phenomenological
 consequences, once the non-locality is implemented in the Standard Model,
 in particular, possible signatures of the NLSM
 at high energy collider experiments.

Let us consider a simple process of 2 by 2 fermion annihilation/creation,
 $f \bar{f} \to f' \bar{f'}$, mediated by photon and $Z$-boson in the $s$-channel.
For a toy version of the NLSM, we apply the modification found
 in the previous section to the vertices and propagators for photon and $Z$-boson.
In a high energy process where fermion masses are approximately taken to be zero,
 the gauge interaction vertex given by (\ref{gauge-vertex}) for a tree-level process remains the same as the Standard Model since the fermion momenta are always on-shell and therefore $k_i^2/M_\mtf^2=m_i^2/M_\mtf^2\approx 0$.
The photon propagator, however,  is modified to have an extra suppression factor $\exp[-s/M^2]$,
 as shown in (\ref{NL-propagator}).
In our toy model approach, we adopt the suppression factor to both photon and $Z$-boson propagators
 with a common nonlocal scale $M$.
As a result, the corresponding scattering cross section in the nonlocal (toy) Standard Model is given by
\ba
 \sigma_{\rm NLSM}(f \overline{f} \to f' \overline{f'})
 = e^{-\frac{s}{M^2}} \times
 \sigma_{\rm SM}(f \overline{f} \to f' \overline{f'}),
\ea
 where $\sigma_{\rm SM}$ is the Standard Model cross section.
 The effect of non-locality is encoded in this ``form factor'' $e^{-s/M^2}$.\footnote{
Similar effects appear in extensions of the Standard Model in the context of non-commutative geometry or TeV scale string theory.
Our results in this section are found to be similar to those in, for example,~\cite{string_LHC, NC_LHC}.
}
For the readers convenience, we list formulas used in the calculation of $\sigma_{\rm SM}$ in the Appendix.

In the following analysis, we make our discussion general and also consider a form factor $e^{+s/M^2}$ with the opposite sign for $M^2$.
We have investigated quantum corrections in nonlocal scalar field theory in Sec.~\ref{sec:scalar}, where loop integrals lead to the exponential-integral
 function $Ei(-m^2/M^2)$.
This function is finite for $m^2/M^2  > 0$ and mildly diverges for $m^2/M^2 \to 0$.
In general, we can change the sign, $M^2 \to -M^2$.
Clearly, in this case, integrand grows exponentially toward high energy, and we need to regularize the loop integral.
However, note that the loop integral leads to $Ei(m^2/M^2)$ and this function is also finite with a mild divergence
 for $m^2/M^2 \to 0$, when we take the the Cauchy Principal Value.
Hence, as long as we follow the Cauchy Principal Value prescription (which we regard as part of the regularization scheme),
 quantum corrections are controlled by the nonlocal scale $M$, and the effect of non-locality can be revealed
 in high energy collider experiments through the form factor $e^{+s/M^2}$.

\begin{figure}[htbp]
\begin{center}
\includegraphics[width=0.7\textwidth, angle=0, scale=1]{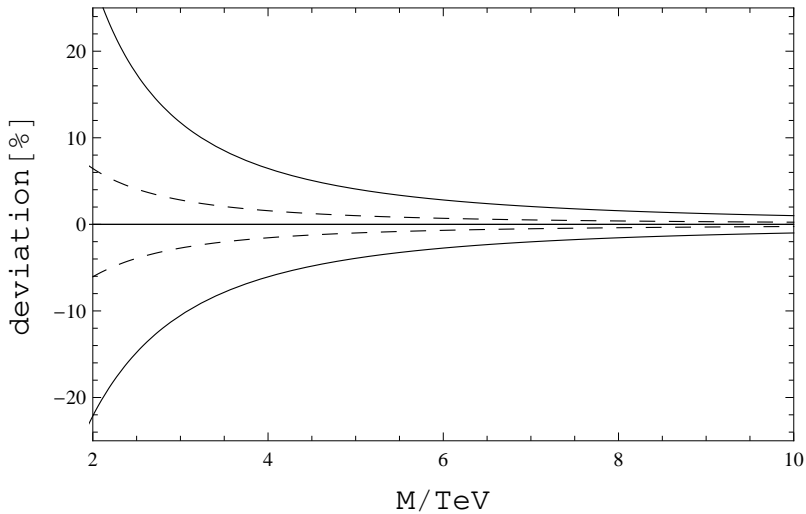}
\end{center}
\caption{
Deviations of the production cross section in the nonlocal Standard Model
 from the Standard Model expectation, as a function of the nonlocal scale $M$
 at the ILC with $\sqrt{s}=500$ GeV (dashed lines) and $1$ TeV (solid lines).
The sign of the deviations corresponds to the form factor $e^{\pm s/M^2}$.
}
\label{fig:ILC}
\end{figure}

We first derive a lower bound on the nonlocal scale $M$ by the results of the LEP experiments.
The cross section of the process $e^+ e^- \to q \bar{q}$ ($q=u,d,c,s,b$) is very precisely measured
 at the LEP experiments with $\sqrt{s}=189$ GeV.
We refer to the results by the OPAL collaboration~\cite{OPAL},
 where the measured cross section is consistent with the Standard Model prediction within a $1.35$\% error.
Since the nonlocal effect must be within this error, we find
\be
 M \geq \frac{189}{\sqrt{-\ln(0.9865)}}~{\rm GeV} \simeq 1.62~{\rm TeV}
\ee
for the form factor $e^{-s/M^2}$, while
\be
 M \geq \frac{189}{\sqrt{\ln(1.0135)}}~{\rm GeV} \simeq 1.63~{\rm TeV}
\ee
for the form factor $e^{+s/M^2}$.
These lower bounds will be increased (or the effect of the non-locality can be discovered) at future collider experiments.
For example, we show in Fig.~\ref{fig:ILC} deviations of the total cross section from the Standard Model expectation,
 $\sigma_{\rm NLSM}/\sigma_{\rm SM}-1$, for the International Linear $e^+ e^-$ Collider (ILC)
 with $\sqrt{s}=0.5$ GeV (dashed line) and $1$ TeV (solid line), as a function of the nonlocal scale $M$.
The ILC, with its high precision, can allow us to test the effect of non-locality
 up to $M \sim 10$ TeV, assuming a 1\% level of precision for the ILC experiments.

We can also investigate the LHC phenomenology, and consider the Drell-Yan process $ pp \to Z/\gamma^* \to e^+ e^-+X$.
The cross section of this process at the parton level, $q \bar{q} \to Z/\gamma^* \to e^+ e^-$ ($q=u, d, s$),
  is enhanced/suppressed by the factor $e^{\pm\frac{s}{M^2}}$ in the toy Standard Model:
 $\sigma_{\rm NLSM}(q {\bar q} \to e^+ e^-; s)=\exp[\pm s/M^2] \times  \sigma_{\rm SM}(q {\bar q} \to e^+ e^-; s)$.
At the LHC, the differential cross section of the process as a function of the $e^+ e^-$ invariant mass is therefore going to be given by~\cite{Schroeder-Peskin}
\begin{eqnarray}
\frac{d \sigma_{\rm NLSM}(pp \to e^+ e^-)}{d M_{ee}} &=&
2 \sum_{q=u, d, s}
\int^{1}_{\frac{M_{ee}^2}{E_{\rm CMS}^2}} d x_1 \frac{2 M_{ee}^2}{x_1 E_{\rm CMS}^2}
f_q \left( x_1, Q ^2 \right) f_{\bar q} \left(  \frac{2 M_{ee}^2}{x_1 E_{\rm CMS}^2}, Q^2 \right)
\nonumber \\
&\times& \sigma_{\rm NLSM}(q \bar{q} \to e^+ e^- ; {\hat s}=M_{ee}^2),
\end{eqnarray}
where $E_{\rm CMS}$ is the collider energy, $M_{ee}$ is the invariant mass of the final state $e^+ e^-$,
 and $f_q$ denotes the parton distribution function of a quark $q$ with the factorizations scale $Q$.
Employing CTEQ5M~\cite{CTEQ5} for the parton distribution functions
 with $Q = M_{ee}$, we calculate the differential cross section for $E_{\rm CMS}=8$ TeV.
Then, we define the deviation from the Standard Model prediction as
\begin{eqnarray}
{\rm deviation}= \frac{d \sigma_{\rm NLSM}}{d M_{ee}}\Big/\frac{d \sigma_{\rm SM}}{d M_{ee}}-1.
\end{eqnarray}

\begin{figure}[htbp]
\begin{center}
\includegraphics[width=0.7\textwidth, angle=0, scale=1]{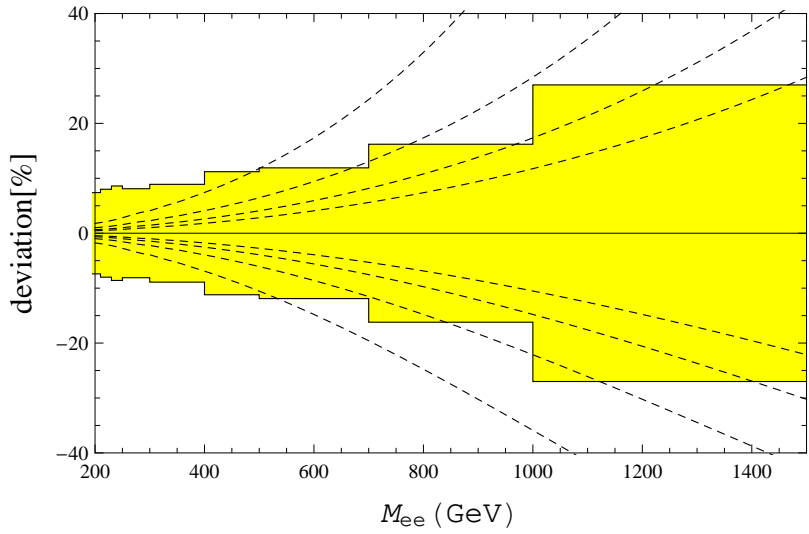}
\end{center}
\caption{
Deviations of the differential cross section in the nonlocal Standard Model
 from the Standard Model expectation for various values of $M$,
The uncertainties of measurements at the ATLAS experiments, including a 2.8 \% uncertainty for the luminisity~\cite{ATLAS_DY},
  are depicted as the shaded region (in yellow).
The sign of the deviations corresponds to the form factor $e^{\pm s/M^2}$.
The dashed lines corresponds to $M=1.5$, $2$, $2.5$ and $3$ TeV, respectively,
 from top in the positive region, while from bottom in the negative region.
}
\label{fig:LHC}
\end{figure}

At the LHC with $\sqrt{s}=8$ TeV, the differential cross section for the $e^+ e^-$ production process
 has been measured by the ATLAS experiment.
The data with a total integrated luminosity of $21.7$/fb have been analyzed, and the results are found to be consistent
  with the Standard Model expectations \cite{ATLAS_DY}.
The uncertainties of the measurement for various $M_{ee}$ values are listed in Table A.12 of Ref.~\cite{ATLAS_DY}.
To find the LHC bound on the nonlocal scale $M$, we refer the results by the ATLAS experiments
 and require the deviation from the Standard Model predictions by the non-locality to be within the uncertainties.
Our results for various values of $M$ are shown in Fig.~\ref{fig:LHC} along with the measurement uncertainties (shaded in yellow).
The dashed lines in the positive region correspond to $M=1.5$, $2$, $2.5$ and $3$ TeV, respectively, from top,
 for the form factor $e^{+s/M^2}$.
The ATLAS results set the lower bound on $M \gtrsim 3$ TeV.
The dashed lines in the negative region correspond to $M=1.5$, $2$, $2.5$ and $3$ TeV, respectively, from bottom,
 for the form factor $e^{-s/M^2}$.
In this case, we find the lower bound on $M \gtrsim 2.5$ TeV.
Note that the results shown as the dashed lines are independent of the LHC energy, but the deviation becomes larger
 as the invariant mass $M_{ee}$ is increasing.
Future LHC experiments with $\sqrt{s}=13-14$ TeV will provide us with the data for higher $M_{ee}$ values,
 leading to more severe constraints (or possible signatures) on the nonlocal scale $M$.

\section{Concluding Remarks}~ \label{sec:conclusions}
In this paper we have taken the first few steps towards constructing a Nonlocal Standard Model where the nonlocality resides in the form of an infinite series of higher derivative terms that do not introduce any new degrees of freedom but exponentially damps UV fluctuations. The modifications are similar to what arises in non-commutative geometry, but they preserve Lorentz invariance, unlike noncommutative field theories. Our considerations were inspired by String theory where such nonlocal operators appear rather frequently such as in String Field theory and p-adic string theories. Now, typically one expect stringy nonlocalities to show up at much larger scales, but there are models where the string scale is $ \cO(\mt{TeV})$ and within reach of the collider experiments. A particularly nice feature of the nonlocal models is to be able to suppress quantum corrections to the mass of the scalar field beyond the scale of nonlocality. This gives us a special incentive to consider a low scale of nonlocality $\sim \cO(\mt{TeV})$ as that would then resolve the hierarchy problem. In this respect the NLSM can be considered as an alternative to the supersymmetric extensions of SM.

Specifically, in this paper we first discussed a nonlocal scalar field theory model. We were able to demonstrate (i) the finiteness of quantum loops in these theories, (ii) the fact that we recover local field theories when the (mass) scale of nonlocality is taken to infinity, (iii) that there appears small corrections in the scattering crosssections as compared to the local field theory results, and finally (iv) that the corrections to the scalar masses indeed grow linearly with the mass scale of nonlocality and therefore one indeed has the potential to address the hierarchy problem within the nonlocal framework.

Next, we implemented nonlocal physics in abelian gauge theories involving massless fermions. We were able to obtain the appropriate Feynman rules for the vertices and propagators in these theories. In particular, for tree-level processes involving onshell massless fermions, there were no nonlocal corrections to the vertex but the propagators received an exponential suppression. We then estimated how the modifications impacts collider phenomenlogy involving two fermion annihilation/creation processes. The LEP data on $e^+e^-\ra q\bar{q}$ scattering and the current LHC data on $q\bar{q}\ra e^+e^-$ gave us bounds on the scale of nonlocality to be around $\sim 1.5$ TeV and $\sim 3$ TeV respectively. We also estimated that with future data coming from the 14 TeV LHC run or a possible ILC run will allow us to probe deviations from SM up to a nonlocality scale of around 10 TeV.

Our construction of the NLSM is however incomplete. For instance, there is no reason to assume that only the abelian sector is nonlocal, in fact, the photon is really a linear combination of the unbroken $SU(2)$ and $U(1)$ gauge fields. Thus it is imperative to be able to generalize the nonlocal interaction to non-abelian gauge theories and implement the Higgs mechanism. One would also like to study how 1-loop processes play out in nonlocal theories. A first step towards this would be to study the $U(1)\times U(1)\ra U(1)$ symmetry breaking with Higgs field. Although this is only a toy model, such investigations should enable one to understand how nonlocal physics impacts Higgs phenomenology.  Eventually the goal will be to construct a fully nonlocal version of SM including the non-abelian gauge sector and  revisit the cross section calculations incorporating $W^{\pm},Z$ mediated Feynman diagrams. In brief, given the current developments in collider experiments nonlocal Standard Models provides an exciting opportunity to explore and test new physics beyond the Standard Model which can have fundamental implications in terms of our understanding of quantum field theories.\vs
{\bf Acknowledgments:} TB would like to thank Joseph Kapusta and Abraham Reddy for various discussions relevant to the project. Also, TB's research was supported in part by the Marquette fellowship grant at Loyola University.

\newpage
\noindent{\Large \bf Appendix}
\appendix
\section{Helicity amplitudes}\label{hel-amp}
Here we provide formulas useful for calculations of the Standard Model cross section,
 $\sigma_{\rm SM} (f\bar{f}  \to f' \bar{f'})$.
We begin with a general formula interaction between a massive gauge boson ($A_\mu$)
 with mass $m_A$ and a pair of the SM fermions,
\begin{eqnarray}
 {\cal L}_{\rm int}= J^\mu A_{\mu}
 =\bar{\psi}_f \gamma^\mu(g_L^f P_L+g_R^fP_R)\psi_f A_{\mu} .
\end{eqnarray}
A helicity amplitude for the process $f(\alpha)\bar{f}(\beta)\rightarrow f'(\delta) \bar{f'}(\gamma)$
is given by
\begin{eqnarray}
 {\cal M}(\alpha,\beta;\gamma,\delta)=
 \frac{g_{\mu \nu}}{s - m_A^2 + i m_A \Gamma_A}
J_{\rm in}^\mu(\alpha, \beta) J_{\rm out}^\nu(\gamma, \delta),
\end{eqnarray}
where $\alpha, \beta$ ($\gamma,\delta$) denote initial (final)
spin states for fermion and anti-fermion, respectively, and
$\Gamma_A $ is the total decay width of the $A$ boson.
We have used 't Hooft-Feynman gauge for the gauge boson propagator
 and there is no contribution from Nambu-Goldstone modes
 in the process with the massless initial (final) states.
The currents for massless initial and final states
are explicitly given by
\begin{eqnarray}
&& J_{\rm in}^\mu(+,-)=-\sqrt{s}g_R^f(0,1,i,0) , \;
      J_{\rm in}^\mu(-,+)=-\sqrt{s}g_L^f(0,1,-i,0) ,
\nonumber \\
&& J_{\rm out}^{\mu}(+,-)= -\sqrt{s} g_R^{f'}(0,\cos\theta,-i,-\sin\theta),
\nonumber \\
&&  J_{\rm out}^{\mu}(-,+)= \sqrt{s} g_L^{f'}(0,-\cos\theta,-i,\sin\theta),
\end{eqnarray}
where
 $\theta$ is the scattering angle, and the other helicity combinations are zero.

The couplings of the Standard Model $Z$ boson with fermions are as follows:
\begin{eqnarray}
 g_L^\nu&=&{e \over \cos\theta_W\sin\theta_W}{1 \over 2}, \; g_R^\nu=0,
\nonumber \\
 g_L^l&=&{e \over \cos\theta_W\sin\theta_W}
   \left(-{1 \over 2}-\sin^2\theta_W(-1)\right), \;  g_R^l=-e(-1)\tan\theta_W,
\nonumber \\
 g_L^u&=&{e \over \cos\theta_W\sin\theta_W}
   \left({1 \over 2}-\sin^2\theta_W{2 \over 3}\right), ];
 g_R^u=-e{2 \over 3}\tan\theta_W,
\nonumber \\
 g_L^d&=&{e \over \cos\theta_W\sin\theta_W}
   \left(-{1 \over 2}-\sin^2\theta_W\left(-{1 \over 3}\right)\right),\;
 g_R^d=-e\left(-{1 \over 3}\right)\tan\theta_W,
\end{eqnarray}
where $e$ is the QED coupling, and $\theta_W$ is the weak mixing angle.
In this paper, we have used the Z-boson mass ($m_Z=91.2$ GeV)
  and its total decay width $\Gamma_Z=2.45$ GeV.
The couplings for the photon are
\begin{eqnarray}
g_L^\nu = g_R^\nu =0, \;  g_L^l = g_R^l = - e, \;
g_L^u   = g_R^u   = \frac{2}{3} e, \;  g_L^d = g_R^d = -\frac{1}{3} e
\end{eqnarray}
with the QED coupling, $e^2/(4 \pi)=1/128$.

\bibliography{fieldtheory-refs}
\bibliographystyle{ieeetr}
\end{document}